\documentclass[prl,aps,floats,twocolumn]{revtex4}

\usepackage[dvips]{graphicx}
\usepackage{amssymb} 
\usepackage{amsmath}
\usepackage{ctable}

\begin{document}

\title{Two-Sheeted Universe, Analyticity and the Arrow of Time}

\author{Latham Boyle$^1$ and Neil Turok$^{1,2}$} 

\affiliation{$^{1}$Perimeter Institute for Theoretical Physics, Waterloo, Ontario, Canada, N2L 2Y5 \\
$^{2}$Higgs Centre for Theoretical Physics, University of Edinburgh, Edinburgh, Scotland, EH8 9YL}

\date{August 2021}
\begin{abstract}
Our universe seems to be radiation dominated at early times, and vacuum energy dominated at late times.  When we consider the maximal analytic extension of this spacetime, its symmetries and complex analytic properties suggest a picture in which spacetime has two sheets, exchanged by an isometry which, in turn, 
picks a preferred (CPT-symmetric) vacuum state for quantum fields on the spacetime.  Previously \cite{Boyle:2018tzc, Boyle:2018rgh}, we showed how this line of thought provides new explanations for dark matter, the matter-antimatter asymmetry, the absence of primordial vector and tensor perturbations, and the {\it phase} of the primordial scalar perturbations; and additional testable predictions.  In this paper, we develop this picture in several respects and, in particular, point out that it also provides a new explanation for why the thermodynamic arrow of time points away from the bang.
\end{abstract}

\maketitle

{\bf Introduction.} Observations reveal that, seconds after the Big Bang, the universe was described by a highly ordered state: a spatially-flat radiation-dominated FRW metric (plus tiny gaussian, adiabatic, purely-growing-mode scalar perturbations described by a nearly-scale-invariant power spectrum) \cite{Ade:2015xua}.  What is this profound clue trying to tell us?  The conventional interpretation is that the radiation dominated era was preceded by an earlier hypothetical epoch of accelerated expansion called inflation.  However, for every property of the universe that inflation claims to explain or predict, there are numerous published inflationary models that claim to explain or predict the opposite.  We worry that inflation may not be a ``good explanation" in the sense articulated in Ch.~1 of \cite{Deutsch}, and hope to find an alternative explanation that is harder to vary.

In Refs.~\cite{Boyle:2018tzc, Boyle:2018rgh}, we began to explore a different approach. Instead of imagining that the simple radiation-dominated era we see was preceded by earlier unseen chaos that must be cleaned up by inflation, we tried to take the observed simplicity of the early universe at face value and suppose that it extends all the way back to the bang, to explore where that line of thought naturally leads.  We found that the maximally-extended spacetime so obtained has a new reflection isometry around the bang, and that taking this symmetry seriously yields elegant new explanations and predictions for several observed properties of the early universe, for the dark matter, and for the neutrino sector.

In this paper, we go further: in addition to being radiation-dominated in the past, our universe will be vacuum energy dominated in the future.   We start by studying the simplest cosmology of this type -- a universe with just radiation and vacuum energy -- paying special attention to the structure (symmetries and complex-analytical properties) of its maximal analytic extension, which we regard as key clues.  We then add matter, requiring it to respect this structure.  We point out that, in addition to the features and predictions pointed out in \cite{Boyle:2018tzc, Boyle:2018rgh}, this picture provides other insights, including a new solution to an old problem: why the thermodynamic arrow of time points away from the bang \cite{PenroseWCH, PenroseR2R}.

\begin{figure}
  \begin{center}
    \includegraphics[width=2.8in]{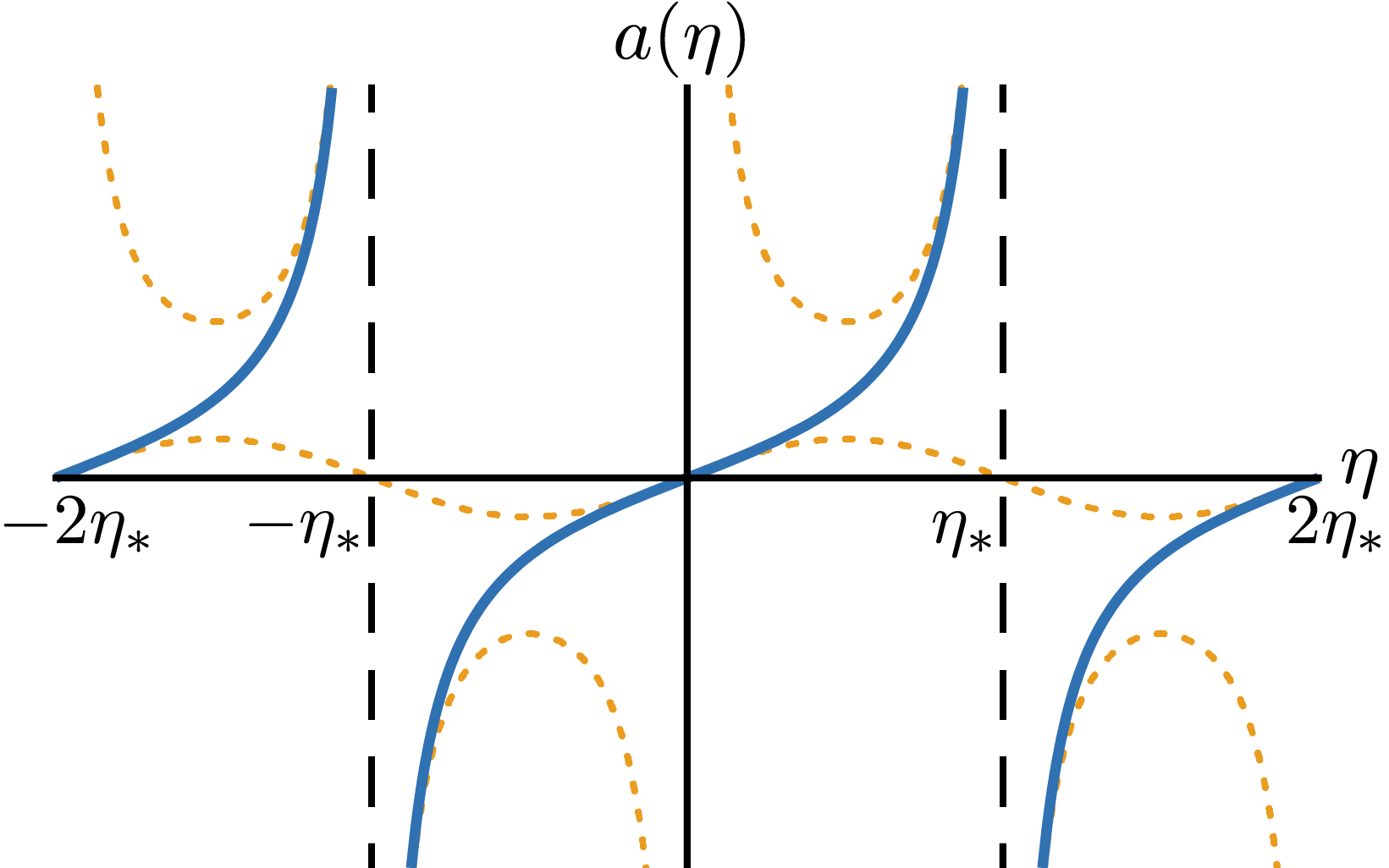}
  \end{center}
  \caption{The real solutions $a(\eta)$.  The solid (blue) curve shows the solution for $3\kappa<\sqrt{4\lambda r}$ (the physically relevant case); and the dashed (orange) curves show them for $3\kappa>\sqrt{4\lambda r}$.}
\label{ScaleFactor}
\end{figure}

{\bf The background spacetime.}  Consider the general FRW line element
\begin{equation}
  ds^{2}=a^{2}(\eta)(-d\eta^{2}+\gamma_{ij}dx^{i}dx^{j})
\end{equation}
where $\gamma_{ij}$ is the metric on a maximally-symmetric space of curvature $\kappa$.  The Friedmann equation is
\begin{equation}
  \frac{\dot{a}^{2}}{a^{4}}=\frac{\rho}{3}-\frac{\kappa}{a^{2}}
\end{equation}
where we have taken $\hbar=c=8\pi G=1$ and a dot denotes $d/d\eta$: $\dot{a}=da/d\eta$.  

Since we would like to study the analytic behavior of the scale factor as $a\to0$ (where radiation dominates) and as $a\to\infty$ (where vacuum energy dominates), we start with the simplest model that has the correct asymptotic behavior at both extremes -- {\it i.e.}\ a universe which only contains radiation and vacuum energy: $\rho=r/a^{4}+\lambda$, where $r$ and $\lambda$ are positive constants.  

Now the Friedmann equation reads:
\begin{equation}
  \label{Friedmann}
  \dot{a}^2={1\over 3} (r-3\kappa a^2 +\lambda a^4).
\end{equation}
Note that in the absence of radiation, the three options for $\kappa$ (positive, negative, and zero) just repesented 3 different ways to slice the same de Sitter (``dS") solution; but when we include radiation, these three options for $\kappa$ represent three physically distinct solutions.

Eq. (\ref{Friedmann}) exhibits an interesting duality: setting $a=1/b$, we find
\begin{equation}
  \label{Friedmanndual}
  \dot{b}^2={1\over 3} (r b^4-3\kappa b^2 +\lambda),
\end{equation}
{\it i.e.}, (\ref{Friedmann}), with $a \rightarrow b$ and $r \leftrightarrow \lambda$. 

\begin{figure}
  \begin{center}
    \hspace*{-4mm}
    \includegraphics[width=2.4in]{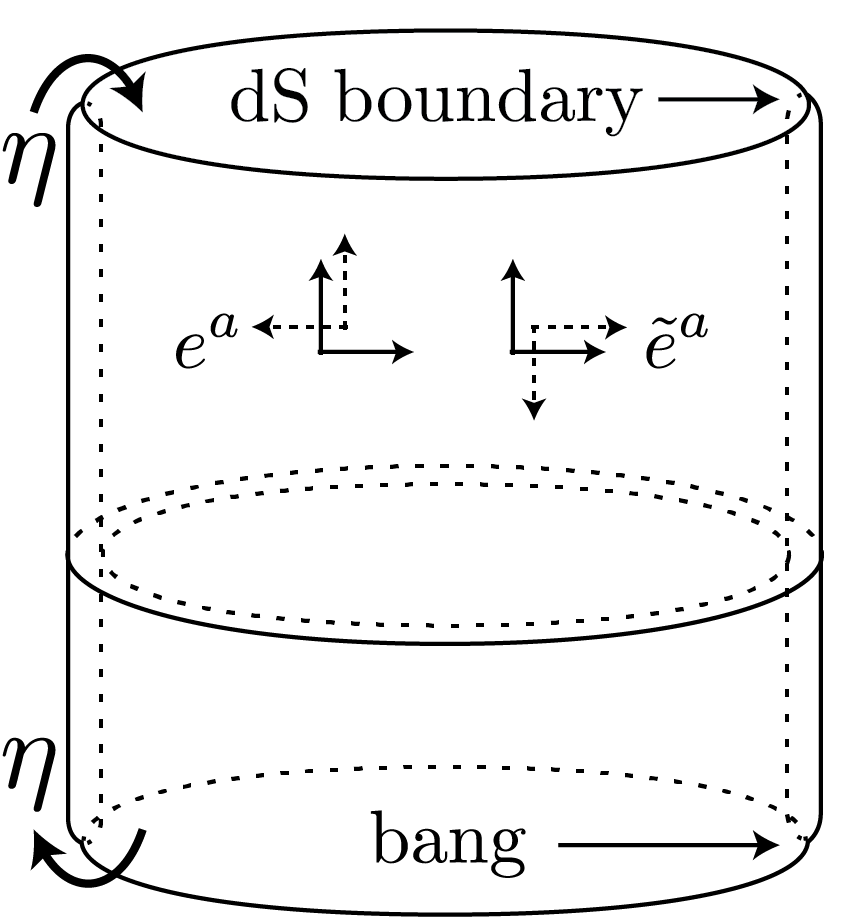}
  \end{center}
  \caption{The two-sheeted universe. The conformal time $\eta$ is circular: as $\eta$ increases, we move up the outer ($a>0$) sheet from the $a=0$ to $a=+\infty$, then back down the inner ($a<0$) sheet from $a=-\infty$ to $a=0$, and so on in a loop.   The metric $g_{\mu\nu}$ and tetrad $e^{a}$ are meromorphic, with a zero
  at the bang, and a pole at dS $\infty$, and the map exchanging each point on the outer sheet with its neighbor on the inner sheet sends $a\to-a$ and is an isometry.
  The conformal metric $\widetilde{g}_{\mu\nu}=a^{-2}g_{\mu\nu}$ and tetrad $\tilde{e}^{\;\!a}=a^{-1}e^{a}$ are $\eta$-independent and non-singular.}
\label{cylinder}
\end{figure}

The solution of (\ref{Friedmann}) is
\begin{equation}
  \label{Friedmann_soln}
 a(\eta)=\alpha\,{\rm sn}(\beta\eta,\,m), 
\end{equation}
where ${\rm sn}(z,m)$ is a Jacobi elliptic function \cite{AbramowitzAndStegun}, $m$ is either root of the quadratic equation
\begin{equation}
  \label{m_eq}
  \frac{(1+m)^2}{m}=\frac{(3\kappa)^{2}}{\lambda r},
\end{equation}
and the constants $\alpha$ and $\beta$ are given by
\begin{equation}
  \alpha=\sqrt{\frac{r(1+m)}{3\kappa}},\qquad\beta=\sqrt{\frac{\kappa}{(1+m)}}.
\end{equation}
The limit $(3\kappa)^{2}/(\lambda r)\to0$ yields the $\kappa=0$ solution.

The choice of roots $m$, $\alpha$ and $\beta$ is just a convention: for definiteness, when $(3\kappa)^{2}<4\lambda r$ (so $m$ is complex) we choose the root $m$ in the upper half plane; and when $(3\kappa)^{2}>4\lambda r$ (so $m$ is real) we choose the root $m<1$ (and imagine it to have an infinitessimal positive imaginary part).  Then $\alpha$ and $\beta$ are fixed by taking $\sqrt{z}$ to have its standard branch cut along the negative real axis.

\begin{figure}
  \begin{center}
    \hspace*{6mm}
    \includegraphics[width=2.8in]{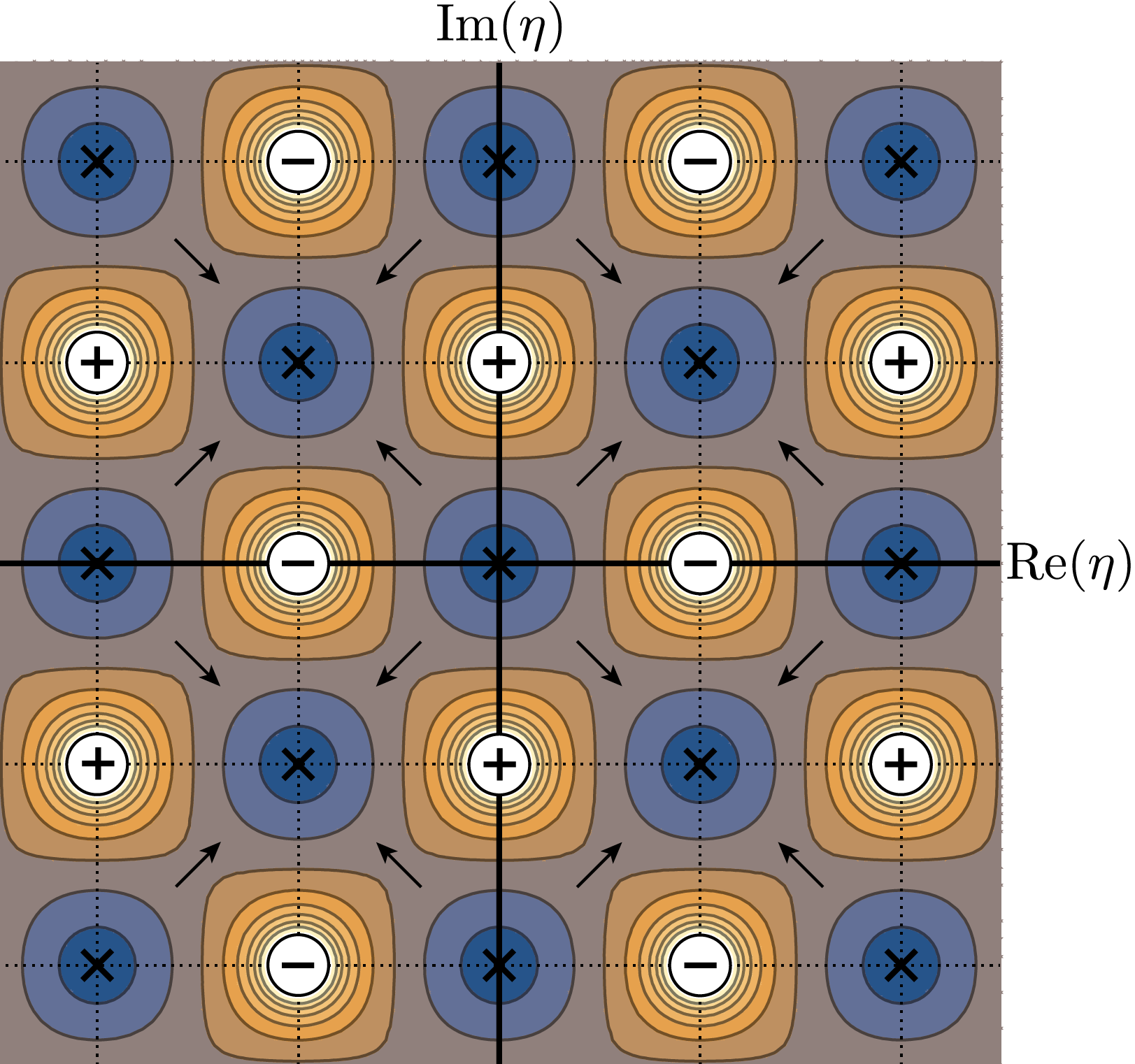}
  \end{center}
  \caption{The complex solution $a(\eta)$, in the complex $\eta$ plane, for the physically relevant case $(3\kappa)^{2}<4\lambda r$ (not too curved).  We have marked
  the zeros with an ``x", and the poles with a ``$+$" or ``$-$" depending on the sign of the residue.  The solid curves show the contours of $|a(\eta)|$; the vertical and   
  horizontal dotted lines show, respectively, where ${\rm Re}[a(\eta)]$ and ${\rm Im}[a(\eta)]$ vanish; and the arrows indicate the direction (phase) of $a(\eta)$ 
  in each of the regions enclosed by dotted lines.}
\label{ComplexPlane}
\end{figure}

Note that ${\rm sn}(z,m)$ is meromorphic in both $z$ and $m$, and doubly periodic in $z$: ${\rm sn}(z,m)={\rm sn}(z+n\omega+n'\omega',m)$, for $n,n'\in\mathbb{Z}$.
The two periods $\omega$ and $\omega'$ point in different directions in the complex plane ({\it i.e.}\ $\omega/\omega'\notin\mathbb{R}$) and are given by $\omega=4K$ and $\omega'=2i K'$, where $K\equiv K(m)$, $K'\equiv K(1-m)$, with $K(m)$ the complete elliptic integral of the first kind \cite{AbramowitzAndStegun}.  Thus, we can think of ${\rm sn}(z,m)$ as a periodic decoration or tiling of the complex $z$ plane, whose fundamental parallelogram has edges $\omega$ and $\omega'$.  Alternatively, we can think of ${\rm sn}(z,m)$ as a function over the toroidal Riemann surface formed by identifying the opposite edges of this fundamental parallelogram.  The zeros of ${\rm sn}(z,m)$ are at $z=2nK+2n'iK'$ and the poles are at $z=2nK+(2n'+1)iK'$, with $n,n'\in\mathbb{Z}$.  

The nature of the solution $a(\eta)$ depends, first, on the sign of the spatial curvature $\kappa$, and second, on whether the universe is close to flat, $(3\kappa)^{2}<4\lambda r$, or far from flat, $(3\kappa)^{2}>4\lambda r$.  In all four cases, as we move along the real $\eta$ axis, the solution $a(\eta)$ is also real, and the universe
described by $a>0$ starts at $\eta=0$ and ends at $\eta=\eta_{\ast}$, where $\eta_{\ast}$ is given in the following table:
\begin{center}
\begin{tabular}{c|c|c}
  & $\quad\;\;\kappa < 0\quad\;\;$ & $\kappa > 0$  \\
  \hline
  $(3\kappa)^{2}<4\lambda r\;\;$ & $iK'/\beta$ & $(2K\!-\!iK')/\beta$  \\
  \hline
  $(3\kappa)^{2}>4\lambda r\;\;$ & $iK'/\beta$ & $2K/\beta$ 
\end{tabular}
\end{center}

In the three cases satisfying $(3\kappa)<\sqrt{4\lambda r}$, $a(\eta)$ describes a universe that grows from a simple zero at $\eta=0$ (the bang) to a simple pole at 
$\eta=\eta_{\ast}$ (dS asymptopia).  In the fourth case $(3\kappa)>\sqrt{4\lambda r}$, corresponding to the lower right-hand corner of the table, $a(\tau)$ describes a universe that grows from a simple zero at $\eta=0$ (the bang) to a finite maximum size, before recollapsing to another simple zero at $\eta=\eta_{\ast}$ (the crunch), while $a(\tau-iK'/\beta)$ describes a universe that shrinks from a simple pole at $\eta=0$ (past dS asymptopia) to a minimum size, before re-expanding to another simple pole at $\eta=\eta_{\ast}$ (future dS asymptopia).  These various solutions are shown in Fig.~\ref{ScaleFactor}.

In all cases, the universe described by $a(\eta)>0$ extends from $\eta=0$ to $\eta=\eta_{\ast}$, but the maximal analytic extension of $a(\eta)$ is periodic, with period $2\eta_{\ast}$, and describes a two-sheeted universe, with $a>0$ on one sheet and $a<0$ on the other, and with $a(\eta)$ odd under temporal reflection across either the bang at $\eta=(2n)\eta_{\ast}$ or the dS boundary at $\eta=(2n+1)\eta_{\ast}$ ($n\in\mathbb{Z}$).   We can visualize this as saying that $\eta$ is curled into a circle, with $\eta$ and $\eta+2\eta_{\ast}$ identified, so that the universe forms the two-sheeted cylinder shown in Fig.~\ref{cylinder}, with an isometry that swaps adjacent points on the inner and outer cylinder and sends $a\to-a$.

{\bf Imaginary time and thermality.}  The analytic structure of $a(\eta)$ is shown in Fig.~\ref{ComplexPlane}, for the physically relevant cases where $(3\kappa)^{2}<4\lambda r$ (``close-to-flat").  As evidence we should take this analytic structure seriously, note the physical time $t(\eta)=\int_{0}^{\eta}a(\eta')d\eta'$ is given by
\begin{equation}
  \label{t}
  t=\frac{\alpha}{\beta}m^{-1/2}{\rm ln}\left(\frac{{\rm dn}(\beta\eta,m)-m^{1/2}{\rm cn}(\beta\eta,m)}{{\rm dn}(0,m)-m^{1/2}{\rm cn}(0,m)}\right),
\end{equation}
where $dn(z,m)$ and $cn(z,m)$ are also Jacobi elliptic functions \cite{AbramowitzAndStegun}, and $\eta$ may be complex.  The integral $\int_{0}^{\eta}a(\eta')d\eta'$ depends on how the contour from $\eta'=0$ to $\eta'=\eta$ winds around the poles of $a(\eta')$.  This in turn corresponds to the fact that the answer (\ref{t}) depends
on ${\rm ln}(z)$, which is only defined modulo $2\pi i$, so that $t$ itself is only well defined modulo $\Delta t=2\pi i(\alpha/\beta)m^{-1/2}=2\pi i\sqrt{3/\lambda}$.  But a system whose time coordinate is only defined modulo an imaginary shift $t\to t+\Delta t$ (or, in other words, whose $t$ is curled into a circle of circumference $\Delta t$ in the imaginary time direction) corresponds to a system at finite temperature $T=i/\Delta t$.  In this way, we precisely recover the well-known expression for the de Sitter temperature
\begin{equation}
  T_{dS}=H/2\pi
\end{equation}
where $H=\sqrt{\lambda/3}$ is the asymptotic dS Hubble constant (the reciprocal of the asymptotic dS radius).

Note the generality of this result: $T_{dS}$ is independent of $\eta$, $\kappa$, or the radiation density, and determined by the residue of $a(\eta)$ at its pole, and hence by $\lambda$.

{\bf Matter and the arrow of time.}  Now let us add matter.  For simplicity, we study a massive scalar field $\varphi$ and leave the details of higher-spin fields to future work.
 
The equation of motion for $\varphi$ is
\begin{equation}
  \label{phi_EOM}
  -\Box\varphi+\xi R\varphi+\mu^{2}\varphi=0
\end{equation}
where the mass $\mu$ is a constant.  For concreteness, we consider the conformal coupling $\xi=1/6$, which is most natural \cite{Penrose, Callan:1970ze}, but our conclusions apply for any value of $\xi$, including minimal coupling $\xi=0$. If we define the $\eta$-independent conformal metric $\tilde{g}_{\mu\nu}=a^{-2}\,g_{\mu\nu}$ and corresponding field $\tilde{\varphi}=a\varphi$, Eq.~(\ref{phi_EOM}) becomes
\begin{equation}
  \label{phi_tilde_EOM}
  -\tilde{\Box}\tilde{\varphi}+\frac{1}{6}\tilde{R}\tilde{\varphi}+\tilde{\mu}^{\;\!2}\tilde{\varphi}=0
\end{equation}
where $\tilde{\Box}$ and $\tilde{R}$ are the wave operator and Ricci scalar of $\tilde{g}_{\mu\nu}$, and $\tilde{\mu}\equiv a\mu$ is time-dependent:
$\tilde{\mu}=\tilde{\mu}(\eta)$.  

We have seen that the analytically-extended background metric may be interpreted as a two-sheeted cylinder, with a reflection symmetry (through the bang or through the dS boundary) which swaps adjacent points on the two sheets (see Fig.~\ref{cylinder}).  We regard this as a clue to be taken seriously.  We thus require that the matter fields living on this background share its structure.  In particular, $\tilde{\varphi}$ should satisfy 
\begin{equation}
  \label{translation}
  \tilde{\varphi}(\eta,{\bf x})=\tilde{\varphi}(\eta+2\eta_{\ast},{\bf x})
\end{equation}
({\it i.e.}\ it lives on the cylinder shown in Fig.~\ref{cylinder}) and 
\begin{subequations}
  \label{reflection}
  \begin{eqnarray}
    \label{reflection_thru_bang}
    \tilde{\varphi}(\eta,{\bf x})&=&\pm\tilde{\varphi}(-\eta,{\bf x}) \\
    \label{reflection_thru_infin}
    \tilde{\varphi}(\eta,{\bf x})&=&\pm\tilde{\varphi}(2\eta_{\ast}-\eta,{\bf x}),
  \end{eqnarray}
\end{subequations}
({\it i.e.}\ it is even or odd under reflection through either the bang or dS boundary: we explain how to resolve this sign ambiguity below).  Note that these conditions are not independent: Eqs.~(\ref{translation}) and (\ref{reflection_thru_bang}) imply Eq.~(\ref{reflection_thru_infin}).

To see what the symmetries (\ref{translation}, \ref{reflection}) imply, let us examine the solutions of (\ref{phi_tilde_EOM}).  For our purposes, it is sufficent to solve (\ref{phi_tilde_EOM}) near the bang and the dS boundary, where the gravitational backreaction of $\varphi$ can be neglected.  \footnote{More generally, the solution $\tilde{\varphi}$ is an arbitrary linear combination of separable solutions $\tilde{\varphi}_{k}(\eta)\psi_{k}({\bf x})$, where $\psi_{k}({\bf x})$ is an eigenfunction of 
$\gamma^{ij}\tilde{\nabla}_{i}\tilde{\nabla}_{j}$; and, if we neglect the gravitational backreaction of $\varphi$, so that $a(\eta)$ is given by (\ref{Friedmann_soln}), then (\ref{phi_tilde_EOM}) becomes the Lam\'{e} equation for $\tilde{\varphi}_{k}(\eta)$, which can be solved in terms of Lam\'{e} functions \cite{WhittakerWatson}.}

Near the bang, where radiation dominates, we can neglect the vacuum energy and spatial curvature of the metric, so $a(\eta)=(r/3)^{1/2}\eta$, and $\tilde{\varphi}$ is a linear combination of terms of the form $\tilde{\varphi}_{k}(\eta){\rm exp}(i{\bf k}\cdot{\bf x})$, where 
\begin{equation}
  \label{EOM_near_bang}
  \ddot{\tilde{\varphi}}_{k}+(k^{2}+\frac{r}{3}\eta^{2}\mu^{2})\tilde{\varphi}_{k}=0, 
\end{equation}
with $k^2={\bf k}\cdot{\bf k}$, and the general solution is 
\begin{equation}
  \label{soln_near_bang}
  \tilde{\varphi}_{k}=c_{+}D_{p}(\gamma\eta)+c_{-}D_{p}(-\gamma\eta),
\end{equation}
where $D_{p}(z)$ is the parabolic cylinder function \cite{GradshteynAndRyzhik}, $c_{\pm}$ are constants, $p\equiv-\frac{1}{2}+i\sqrt{3/r}\frac{k^{2}}{2\mu}$, and $\gamma\equiv{\rm e}^{-i\pi/4}(\frac{4r}{3}\mu^{2})^{1/4}$.  

Near the dS boundary, where vacuum energy dominates, we can neglect the radiation and spatial curvature of the metric, so $a(\eta)=1/H(\eta_{\ast}-\eta)$, and $\tilde{\varphi}$ is a linear combination of terms $\tilde{\varphi}_{k}(\eta){\rm exp}(i{\bf k}{\bf x})$, where 
\begin{equation}
  \label{EOM_near_dS}
  \ddot{\tilde{\varphi}}_{k}+(k^{2}+\frac{\mu^{2}}{H^{2}(\eta_{\ast}-\eta)^{2}})\tilde{\varphi}_{k}=0,
\end{equation} 
with the general solution 
\begin{equation}
  \label{soln_near_dS}
  \tilde{\varphi}_{k}=(k(\eta_{\ast}\!-\eta))^{1/2}[c_{1}J_{\nu}(k(\eta_{\ast}\!-\eta))\!+c_{2}Y_{\nu}(k(\eta_{\ast}\!-\eta))]
\end{equation}
where $J_{\nu}(z)$ and $Y_{\nu}(z)$ are Bessel functions \cite{GradshteynAndRyzhik}, $c_{1}$ and $c_{2}$ are constants, and $\nu\equiv\frac{1}{2}(1-4m^{2}/H^{2})^{1/2}$.  

From these solutions, we see a crucial difference between the behavior at the bang and at the dS boundary.  At the bang, $\tilde{\varphi}_{k}$ remains analytic ($\varphi_{k}=a^{-1}\tilde{\varphi}_{k}$ remains meromorphic), and may be analytically continued across it.  For this analytic continuation to be compatible with the reflection symmetry (\ref{reflection}), $\tilde{\varphi}$ must satisfy the following boundary condition (``b.c.") at the bang:
\begin{subequations}
  \label{bcs}
  \begin{eqnarray}
    \tilde{\varphi}(0,{\bf x})&=&\pm\tilde{\varphi}(0,{\bf x}), \\
    \dot{\tilde{\varphi}}(0,{\bf x})&=&\mp\dot{\tilde{\varphi}}(0,{\bf x}).
  \end{eqnarray}
\end{subequations}
By contrast, $\tilde{\varphi}$ is {\it not} analytic at the dS boundary, and cannot be analytically extended across it, so it does {\it not} satisfy a corresponding b.c.\ there.  

{\it The fact that basic considerations of symmetry and analyticity force all fields to satisfy a restrictive boundary condition at one end of spacetime (the bang) but not at the other end (the dS boundary) is striking.  It seems to suggest a fundamental new explanation for why the thermodynamic arrow of time ({\it i.e.}\ the direction in which entropy increases) points away from the bang.} 

The upper (lower) sign in (\ref{bcs}) corresponds to a Neumann (Dirichlet) b.c.\ for $\tilde{\varphi}$ at the bang.  This is closely related to the b.c. at the boundary of anti de Sitter (AdS) spacetime \cite{Avis:1977yn, Breitenlohner:1982jf}: there too, if one conformally maps the spacetime to a portion of the infinite static (Einstein) cylinder, one finds a simple reflecting (Neumann or Dirichlet) b.c.\ for the field $\tilde{\varphi}$.  In fact, the AdS b.c.s were originally derived \cite{Avis:1977yn} by mapping AdS to a portion of the static Einstein cyclinder, and then noting that it could be analytically extended to form a two-sheeted spacetime, with a natural $\mathbb{Z}_{2}$ symmetry relating the sheets.  Of course, this closely parallels our derivation, but with some noteable differences: in the AdS case, one analytically continues in the radial direction, and the resulting b.c.s make the AdS boundary an ordinary mirror (a timelike surface which reflects radially); while in our case, one analytically continues in the time direction, and the resulting b.c.s make the bang an exotic mirror (a spacelike surface which reflects in the time direction).

For a canonical scalar field, it seems most natural to choose the lower sign in (\ref{bcs}) at the bang, so that $\tilde{\varphi}$ and $\varphi=a^{-1}\tilde{\varphi}$ satisfy Dirichlet and Neumann boundary conditions, respectively, since in this case $\tilde{\varphi}$ and $\varphi$ both remain finite and analytic at the bang, consistent with the assumption of small gravitational back reaction. 

The $\pm$ option in (\ref{bcs}) also appears in the b.c.s for higher spin fields (as in the AdS case \cite{Breitenlohner:1982jf, Hawking:1983mx}).   Since $a(\eta)$ is odd under $\eta\to-\eta$, we can infer right choice for the tetrad: the conformal tetrad $\tilde{e}^{\;\!a}_{\mu}$ (with $\eta_{ab}\tilde{e}^{\;\!a}_{\mu}\tilde{e}^{\;\!b}_{\nu}=\tilde{g}_{\mu\nu}$) satisfies a Neumann b.c. (or, equivalently, the symmetric condition $\tilde{e}^{\;\!a}_{\mu}(\eta,{\bf x})=\tilde{e}^{\;\!a}_{\mu}(-\eta,{\bf x})$) while the physical tetrad $e^{a}_{\mu}=a\tilde{e}^{\;\!a}_{\mu}$ (with $\eta_{ab}e^{a}_{\mu}e^{b}_{\nu}=g_{\mu\nu}$) satisfies a Dirichlet b.c. (or, equivalently, the anti-symmetric condition $e^{a}_{\mu}(\eta,{\bf x})=-e^{a}_{\mu}(-\eta,{\bf x})$).  As shown in \cite{Boyle:2018tzc}, this condition successsfully explains: (i) the absence of primordial vector perturbations; (ii) the absence of primordial ``decaying-mode" scalar and tensor perturbations (which synchronizes the scalar perturbations, explaining the observed oscillations in the CMB temperature spectrum and their phases); and (iii) the fact that bang seems to be a particularly mild singularity (a ``Weyl singularity" -- a non-singular conformal metric times an overall conformal factor $a(\eta)$ which momentarily passes through zero).
     
{\bf Remarks.} (i) In a forthcoming paper, we present a mechanism for generating the observed scale-invariant scalar perturbations in this framework~\cite{perts}.

(ii) In our picture, instead of thinking of the universe evolving {\it through} the bang, we can think of the bang as a spacelike mirror -- a surface of reflection symmetry.  As with an electromagnetic mirror, we can either think of the bang as boundary of spacetime, satisfying special boundary conditions, or else (by the method of images), we can think of a mirror universe on the ``other side" of the bang.

(iii) One might worry that, in deriving (\ref{Friedmann_soln}), we dropped the only term in the Friedmann equation (the matter density $\propto a^{-3}$) that is odd under $a\to-a$, while all other terms are even.  Would including this term spoil the symmetry?  No: if the matter fields satisfy the condition (\ref{bcs}) then, far from the bang, the matter density is actually $\propto |a|^{-3}$ (which is even under $a\to-a$); while near the bang it remains perfectly analytic.

(iv) We saw that the solution (\ref{soln_near_dS}) generally has an essential singularity near the dS boundary.  The exception is when the field is massless ($\mu=0$), in which case (\ref{soln_near_dS}) remains analytic at the dS boundary \footnote{The fact that both massless and massive solutions are analytic at the bang, while only massless solutions are analytic at the dS boundary is due, physically, to the fact that both massless and massive particles reach the bang in finite proper time, while only massless particles reach the dS boundary in finite proper time.}.  This leaves open the possibility that massless fields might obey a boundary condition there too, which would quantize the allowed comoving wavenumbers for massless fields~\cite{perts}.   

(v) It is natural to speculate that our classical symmetry conditions $e^{a}_{\mu}(-\eta,{\bf x})=-e^{a}_{\mu}(\eta,{\bf x})$ and $\varphi(-\eta,{\bf x})=\varphi(\eta,{\bf x})$ (along with analogous conditions for the various other fields) follow from a path integral formulation in which only paths satisfying these conditions are allowed \footnote{In this picture, the bang may alternatively be thought of as a kind of spacelike orbifold boundary, with some similarity to the timelike orbifold boundaries which appear in constructions like those in \cite{Horava:1995qa, Horava:1996ma}.}.    This suggests a natural wavefunction of the universe: namely, that the amplitude for a given spacelike 3-configuration of the geometry and matter fields is the path integral over all appropriately symmetric 4-configurations ending on the given 3-configuration.

{\bf Acknowledgements.}  Research at Perimeter Institute is supported by the Government of Canada, through Innovation, Science and Economic Development, Canada
and by the Province of Ontario through the Ministry of Research, Innovation and Science. The work of NT is supported by the STFC Consolidated Grant “Particle Physics at the Higgs Centre.”

\end{document}